\documentclass{elsart}

\usepackage{graphics}
\usepackage{graphicx}

\begin{document}

\begin{frontmatter}



\title{Quantitative AFM analysis of phase separated borosilicate glass surfaces}


\author{D. Dalmas\corauthref{cor1}},
\ead{davy.dalmas@saint-gobain.com}
\corauth[cor1]{corresponding author}
\author{A. Lelarge }and
\author{D. Vandembroucq}

\address{Laboratoire ``Surface du Verre et Interfaces'' \\
Unit\'e Mixte CNRS/Saint-Gobain \\
39 Quai Lucien Lefranc, 93303 Aubervilliers, France}

\begin{abstract}
Phase separated borosilicate glass samples were prepared by applying various
heat treatments.  Using selective chemical etching we performed AFM
measurement on the phase separated glass surfaces. A quantitative
roughness analysis allowed us to measure precisely the dependence of
the characteristic size of the phase domains on heating time and
temperature. The experimental measurements are very well described by
the theoretically expected scaling laws. Interdiffusion coefficients
and activation energy are estimated from this analysis and are
consistent with literature data.
\end{abstract}

\end{frontmatter}


\section{Introduction}
Within the last 20 years Atomic Force Microscopy (AFM) has become a
standard tool in glass science\cite{Arribart-00}. Beyond the direct
roughness measurements, AFM has been applied to study the glass
structure, corrosion, fracture, polishing and progressively more to
characterize various (organic or inorganic) coatings deposited on
glasses. Increasingly precise measurements have been obtained in the
last years allowing to reach atomic resolution
\cite{Raberg-JNCS05,Frischat-JNCS04}.
Apart from detailed studies of the scale invariant character of the
roughness of fracture surfaces\cite{Daguier-PRL97} or the recent
description of fused glass surfaces in terms of frozen capillary
waves\cite{SLSV-Preprint06}, AFM roughness analysis is often restricted
to the simple determination of the standard deviation of the height
fluctuations, {\it i.e.} the RMS roughness. This RMS roughness
indicator which only gives information on the out of plane surface
fluctuations is however a rather poor information as compared to the
available information in the entire image.  The characterization of
in-plane features, {\it e.g.} the typical size of patterns on the
surface requires the use of the height auto-correlation function.

In this paper we develop a quantitative analysis of AFM roughness data
to study the kinetics of phase separation coarsening stage in glasses.
AFM measurement has been used to study phase separation but mostly to
image the surface of thin polymer films
\cite{kim-jms99,zdravkova-jcg05,Cyganik-ss02,rasmont-csb00,ichii-ass03,bergues-v01}. The
development of phase separation in glasses has been intensively
studied experimentally in the last
decades\cite{MazurinBook84}. Various techniques have been used to
follow this phenomenon : visible or X
scattering\cite{tanaka-jncs84,polizzi-jncs98,polizzi-jac97,tomozawa-jacs80,Danforth-jncs83},
electronic microscopy\cite{Elmer-jacs70,polizzi-mmm97}, infrared or
Raman
spectroscopy\cite{kato-jacs01,koni-jncs76,White-jncs82,sigoli-jncs01}...
Most of the times, these techniques are used to test the occurrence of
phase separation and to describe its morphology (droplet or
interconnected). Few studies (generally based on electronic microscopy
or X-ray scattering) give a quantitative estimate of the volume
fraction of the two separated phases and sometimes of the typical
size of the domains. X-ray and visible light scattering techniques
have been intensively used to characterize the kinetics of phase
separation, from the early stage of binodal or spinodal phase
decomposition up to the scaling behavior of the coarsening
stage\cite{MazurinBook84}. More recently, coherent X-ray techniques
have been used to probe the dynamics of fluctuations during domain
coarsening of a sodium borosilicate glass\cite{Malik-PRL98}.

In the present work we focus our study on a sodium-potassium
borosilicate glass. Alkali borosilicate glasses\cite{MazurinBook84}
are known for their large composition domain allowing phase
separation. After heat treatment and etching we perform AFM surface
measurements in order to reveal the internal structure.The paper is
organized as follows. We first describe the experimental methods:
melting, sample preparation, heat treatments, chemical etching and AFM
measurements. We then present the topographic results obtained by AFM,
giving a special attention to the statistical analysis. We finally
compare our results with theoretical descriptions for phase separation
kinetics and we extract estimates of interdiffusion coefficient and
activation energy.

\section{Experimental}
\subsection{Material}
A sodium potassium borosilicate glass of weight composition
$\mathrm{SiO}_{2} \; 70\%$, $\mathrm{B}_{2}\mathrm{O}_{3}$ $25\%$,
$\mathrm{Na}_{2}\mathrm{O}\; 2.5\%$ and $\mathrm{K}_{2}\mathrm{O}\;
2.5\%$ has been prepared as follows: i) The raw materials are mixed
and melted at 1550$^\circ$C in a 800 $\mathrm{cm}^3$ platinum
crucible; ii) The glass is then refined during 2 hours in order to
obtain an homogeneous glass; iii) After quenching, glass is rapidly
transferred to an electric furnace to be annealed during 1h at
630$^\circ$C, in order to relax the internal thermal stresses; iv) The
glass is finally cooled to room temperature by opening the furnace
door. After preparation, a cylindrical sample is core drilled and then
cut into centimeter thick slices.  At that stage, the glass is no
longer transparent. Its opalescence is a signature of phase separation
taking place over length scales comparable to the wavelengths of
visible light. The major phase is expected to have a concentration
close to pure silica while the minor phase concentrates the other
constituents\cite{MazurinBook84}.

\subsection{Thermal treatments}

Two types of thermal treatments have been performed. Four thick slices
are first submitted to a thermal treatment at 650$^\circ$C of duration
1h, 4h, 16h and 64h respectively. Square samples of 0.5mm thickness
are cut from the slices and then mechanically polished to obtain
mirror smooth surfaces. Following this first method, the result of
bulk phase separation can be observed at the surface.  A second sample
preparation has consisted of inverting the sequence: first cutting the
square samples from a glass slice, polishing them and only then
performing the thermal treatments.  Following this second method, we
can observe the influence of a surface on phase separation.  Two
series of thermal treatments have been performed, one at 650$^\circ$C
with seven different durations (1h, 2h, 4h, 8h, 16h, 32h, 64h and 96h)
and a second one at six different temperatures (625$^\circ$C,
650$^\circ$C, 660$^\circ$C, 670$^\circ$C, 675$^\circ$C and
680$^\circ$C) during 16h. The first series allows us to study the
influence of thermal treatment duration on phase separation
morphology. The second one allows investigation of the influence of
temperature. In all cases, the samples are heat-treated in an electric
furnace and cooled to room temperature outside the furnace.

\subsection{Etching}
A selective attack with a slightly acid treatment ($pH\approx5$) is
performed. The borate rich phase is thus eliminated in surface due to
its very low chemical durability while the silica rich phase remains
almost unaffected. The residual surface roughness then makes it
possible to reveal the phase separation morphology as the composition
contrast induces a height contrast. The alkali borate rich phase is
thus expected to appear as depressions compared to the almost
unmodified mirror surface corresponding to the silica rich phase.

\subsection{AFM measurements}
The surface morphology of all samples have been characterized by AFM
height measurement in tapping mode (TM-AFM), using a Nanoscope III A
from Digital Instruments with Al coated tip (BudgetSensors - model
BS-Tap 300 Al). The Al coating thickness is 30nm, the resonant
frequency is 300 kHz and the stiffness constant is 40N/m. The images
have been recorded at a scan frequency between 0.8 and 1Hz for a
resolution of $512\times512$ pixels.  For each samples (i.e. for each
thermal treatment conditions), we perform at least 6 AFM images with 3
different scan areas.  Depending on the size of the phase separation
domains, the scan areas used are $1\times1\mu m^{2}$, $2\times2\mu
m^{2}$, $4\times4\mu m^{2}$ and $8\times8\mu m^{2}$.

\section{Results}
\subsection{Morphology of phase separation domains}

\begin{figure}
\begin{center}
\includegraphics[width=13cm]{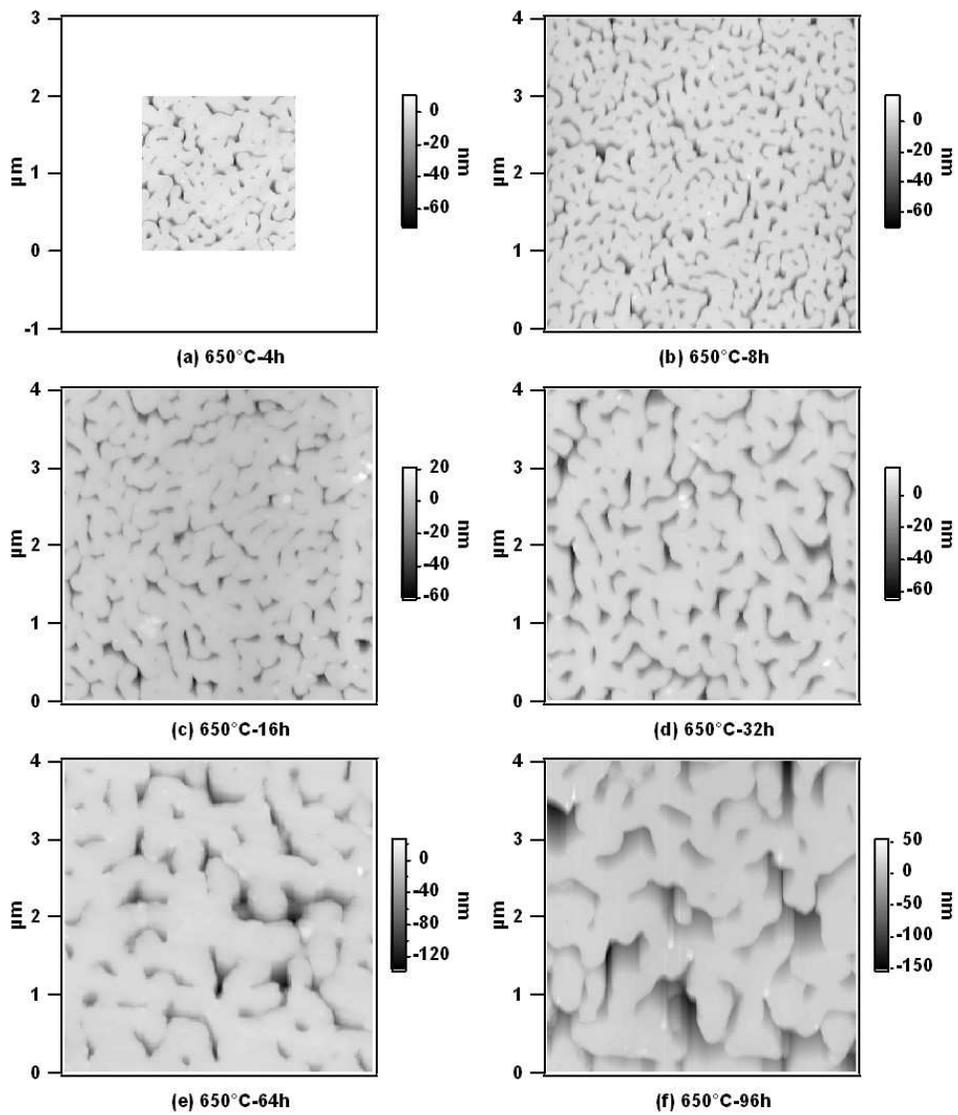}
\caption {AFM height images of borosilicate glass samples after thermal treatment at 650$^\circ$C during  six different durations.}
\label{tps_6AFM}
\end{center}
\end{figure}

\begin{figure}
\begin{center}
\includegraphics[width=13.5cm]{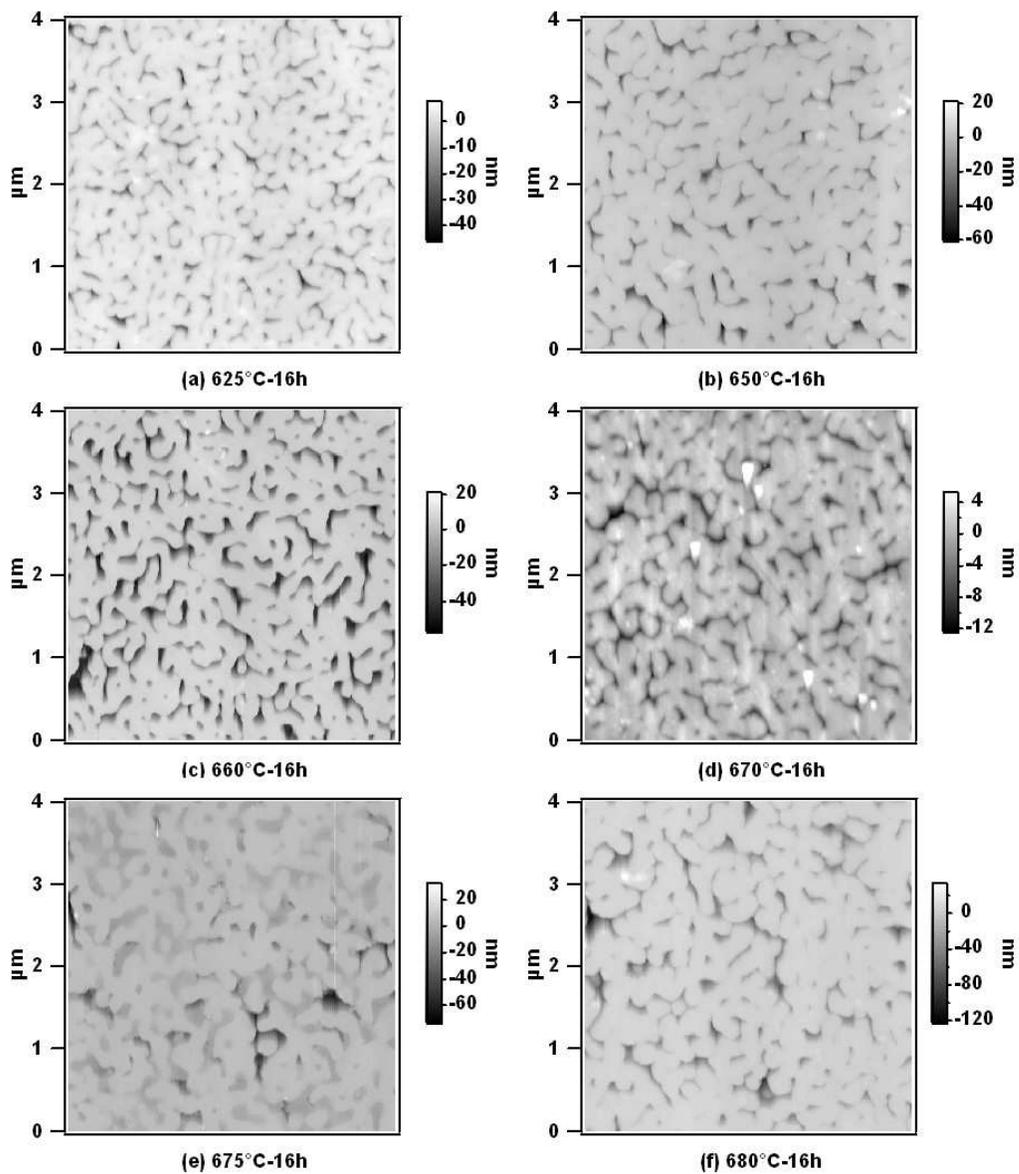}
\caption {AFM height images of borosilicate glass samples after 16 hours thermal treatments at six different temperatures.}
\label{temp_6AFM}
\end{center}
\end{figure}

For all samples (cf. Fig.~\ref{tps_6AFM} and Fig.~\ref{temp_6AFM}),
we observed an interconnected morphology. On all these images, the
grey scale is encoded such that the silica rich phase appears in
clear colors and the borate rich phase appears in dark colors. On
Fig.~\ref{tps_6AFM} (respectively on Fig.~\ref{temp_6AFM}), one can
easily see the influence of time (resp. temperature) of thermal
treatment on the morphology: at a given temperature (resp.
time), the characteristic size of spinodal phase separation
domains increases with annealing time (resp. temperature).

\begin{figure}
\begin{center}
\includegraphics{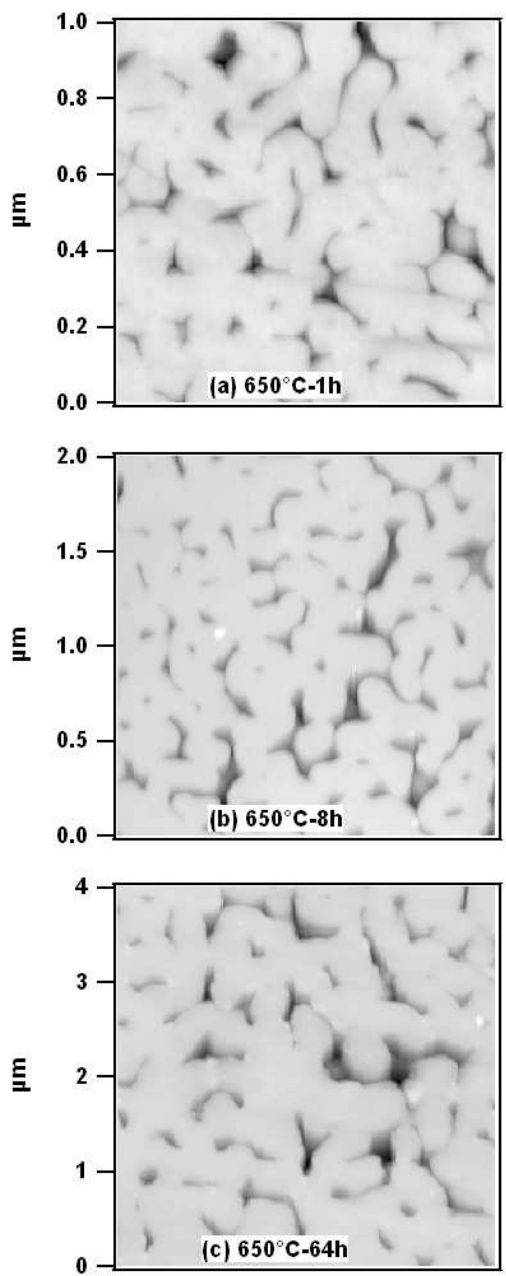}
\caption {AFM height images of three borosilicate glass samples. These images illustrate the similarity of morphology of the phase separation domains for various observation scales according to the duration of thermal treatment at 650$^\circ$C.}
\label{ech}
\end{center}
\end{figure}

As clearly shown on Fig.~\ref{ech}, we obtained patterns exhibiting a
striking self-similar character. In particular with a well chosen set
of thermal treatment duration and AFM image size, we obtain almost
indistinguishable images. This self similar character simply reflects
that space and time can be rescaled altogether. We show below that AFM
measurements allow us to go beyond this qualitative picture. In
particular we will evidence that the kinetics of the phase domains
obeys the expected scaling law $\xi \propto t^{1/3}$ between the
characteristic size $\xi$ and the duration $t$ of the thermal
treatment (at a given temperature) in the coarsening stage of the
phase separation process\cite{MazurinBook84}.

\subsection{Statistical analysis of AFM images}

In the following we use AFM measurements on etched surfaces to
estimate the characteristic size of the phase separated domains. AFM
measurements give an immediate access to the height field. To get
information on the growth of the phase domains we will exploit the
height autocorrelation function.

\subsubsection{Determination of the radial-mean auto-correlation function}
\begin{figure}
\begin{center}
\includegraphics{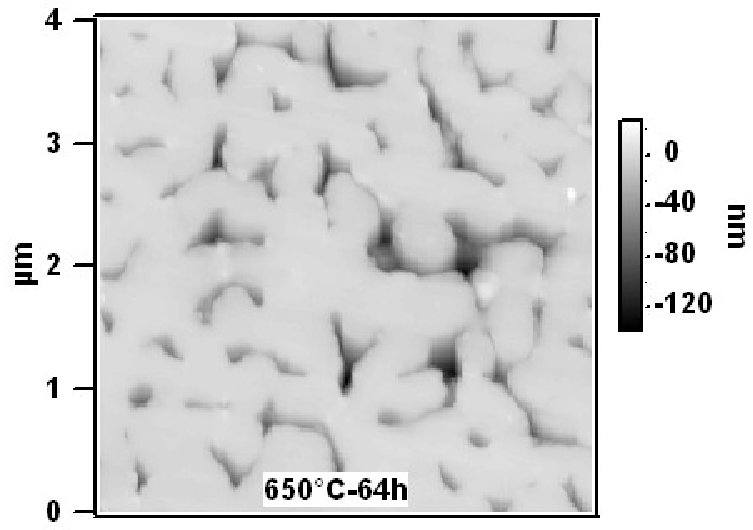}\includegraphics[width=6cm]{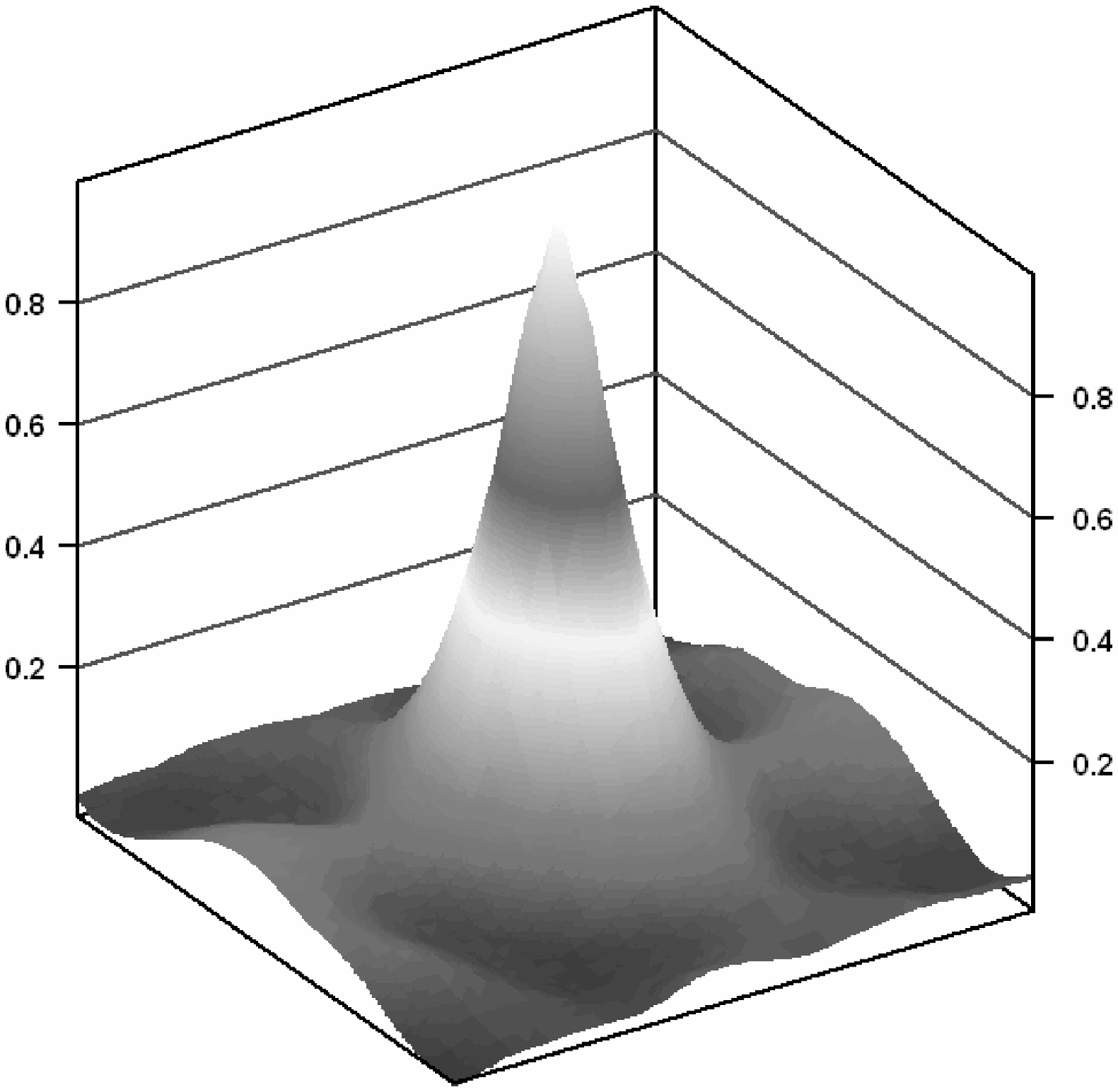}
\caption {AFM height images ($4\times4\mu m^{2}$) of a borosilicate glass sample after 64 hours of thermal treatment at 650$^\circ$C (left) and the associated three dimensional representation of the center part ($0.5\times0.5\mu m^{2}$) of the auto-correlation function $C(\vec{x})$ (right).}
\label{corr}
\end{center}
\end{figure}

\begin{figure}
\begin{center}
\includegraphics{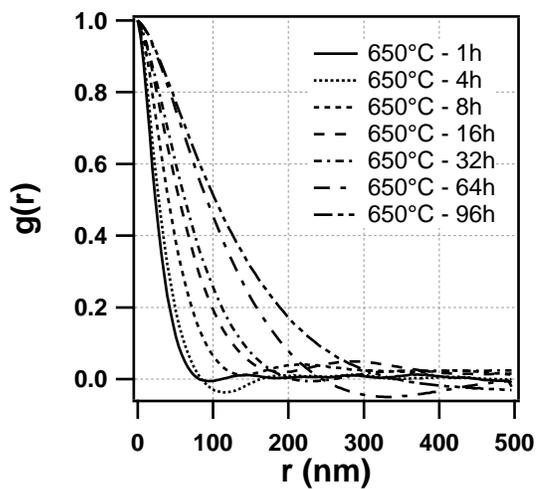}
\caption {Radial mean value of the auto-correlation function $g(r)$ {\it vs}  distance $r$ (in nm) for borosilicate glass samples annealed during different durations at 650$^\circ$C.}
\label{tps_corrlin}
\end{center}
\end{figure}

\begin{figure}
\begin{center}
\includegraphics{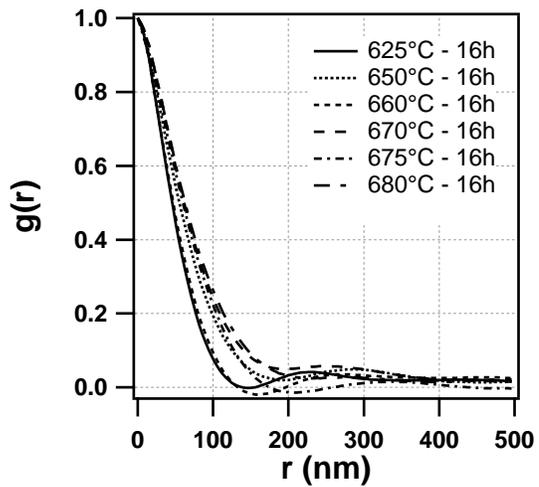}
\caption {Radial mean value of the auto-correlation function $g(r)$ as a function of distance $r$ (in nm) for borosilicate glass samples annealed during 16h at different temperatures.}
\label{temp_corrlin}
\end{center}
\end{figure}

In order to evaluate the characteristic length of phase separation
domains for all our samples, we first calculate for all TM-AFM
height images the normalized auto-correlation function $C(\vec{R})$,
given by the following equation:
\begin{equation}
    C(\vec{R})=\frac{\langle h(\vec{x})\cdot h(\vec{x}+\vec{R})\rangle_{\vec{x}}}{\langle h(\vec{x})\cdot h(\vec{x})\rangle_{\vec{x}}}
\end{equation}
where $h(\vec{x})$ is the height in nanometers at a point of
coordinate $\vec{x}(r,\theta)$ and with $C(\vec{0})=1$. On Fig
~\ref{corr}, we give an example of the three dimensional
representation of the center part ($0.5\times0.5\mu m^{2}$) of the
auto-correlation function $C(\vec{R})$ calculated from the height
matrix extracted from an AFM height images ($4\times4\mu m^{2}$). 
On this particular 3D representation and also for all other
images, we obtain results characteristic of an isotropic random
distribution of domains: a central peak surrounded by a few circular
rings, corresponding to oscillations of vanishing amplitude. The
isotropic character of the fluctuations allows for the use of an average auto-correlation function dependent on the only distance:
\begin{equation}
  g(r)=\langle C(r,\theta)\rangle_{\theta}
\end{equation}
Fig. \ref{surf-bulk} shows the auto-correlation function for a heat
treatment of 16h at 650$^\circ$C performed on two series of samples
where the heat treament has been applied before or after cutting and
polishing operations.  The evolution of $g(r)$ as a function of the
duration of heat treatment at 650$^\circ$C and respectively as a
function of temperature for a 16h heat treatment is given on
Fig.~\ref{tps_corrlin} and Fig.~\ref{temp_corrlin} respectively. On
these two figures, $g(r)$ corresponds to the mean value of the
auto-correlation function of the six AFM images associated with one
sample.

\subsubsection{Determination of two correlation lengths}
\begin{figure}
\begin{center}
\includegraphics[width=10cm]{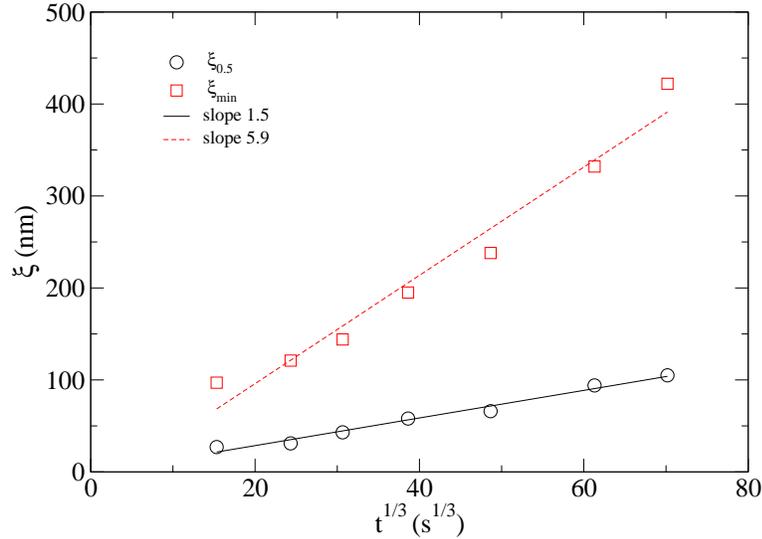}
\caption {Evolution of the two correlation length ($\xi_{0.5}$ and $\xi_{min}$) w.r.t. the duration of thermal treatment at 650$^\circ$C.}
\label{tps_lc}
\end{center}
\end{figure}

\begin{figure}
\begin{center}
\includegraphics[width=10cm]{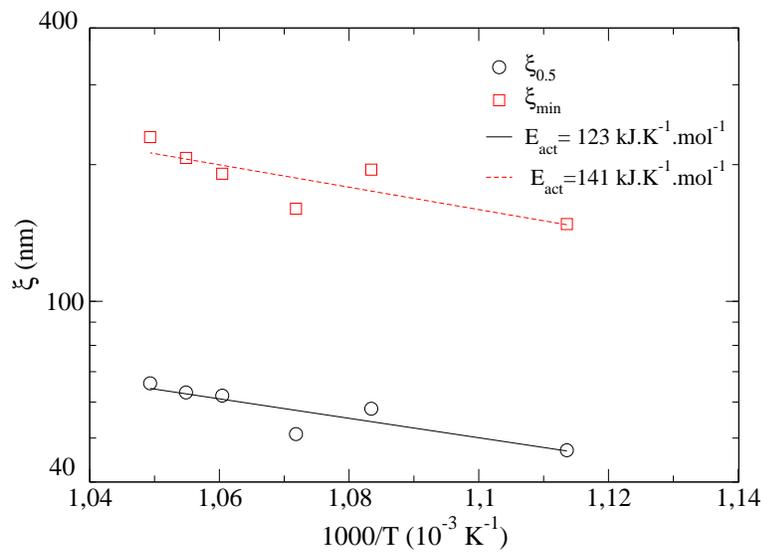}
\caption {Evolution of the two correlation lengths ($\xi_{0.5}$ and $\xi_{min}$) w.r.t. the temperature for 16 hours thermal treatments.}
\label{temp_lc}
\end{center}
\end{figure}

The second step of our analysis of AFM height data consists of
extracting for each sample a characteristic length from auto
correlation function $g(r)$. We discuss here two typical lengths,
chosen for their physical meaning with respect to phase separation
phenomena. The first one, called $\xi_{0.5}$, is the value of $r$ at
$g(r)=0.5$ ($g(\xi_{0.5})=0.5)$) and corresponds to the mean size of
the phase separation domains. The second one, called $\xi_{min}$, is
the value of $r$ when $g(r)$ reached its first minimum
($g(\xi_{min})=min(g(r))$) and corresponds to the mean distance between
two domains of different composition. The evolution of these
characteristic lengths with heat treatment duration and temperature are
summarized in tables ~\ref{tabtps} and ~\ref{tabtemp} and displayed in
Fig.~\ref{tps_lc} and Fig.~\ref{temp_lc} respectively. These
results are discussed in next section in the light of theoretical
predictions for the kinetics of the coarsening stage of phase
separation in glasses.

\section{Discussion}

The process of phase separation in a spinodal region is usually divided
into two stages.  The first one corresponds to the phase decomposition
into domains at the two limit concentrations. The second one is the
coarsening process, corresponding to the progressive growth of the
domains, driven by surface tension and limited by diffusion.
During this coarsening regime, the structure remains statistically
self-similar and the evolution of the morphology is entirely
controlled by the growth of the characteristic length of the domains
which is expected to follow the Lifshitz-Slyozov-Wagner scaling:

\begin{equation}\label{scaling}
  R\approx (Kt)^{1/3}\;,\quad  K \approx \frac{\gamma D v}{kT}\;,
\end{equation}
where, within the classical mean field theory of
coarsening\cite{MazurinBook84,BalluffiBook05} $\gamma$ is the
interfacial tension, $D$ the interdiffusion coefficient and $v$ a
molecular volume. This result applies in case of conservation of the
order parameter. Following Huse\cite{Huse-PRB86}, it can be recovered
{\it via} a simple scaling argument. Let us consider an interface of
radius of curvature $R$ separating two domains with a volume
concentration difference $\Delta c$. Under local equilibrium, the
Laplace pressure $\gamma/R$ balances the osmotic pressure $\mu \Delta
c$ (where $\mu$ is the chemical potential) so that $\mu \simeq \gamma
/R\Delta c$. Assuming that fluctuations of the chemical potential are
of the same order of magnitude as the chemical potential itself we
have $\nabla \mu \simeq \gamma /R^2\Delta c$. The flux of particles
can thus be written $j=Mc\nabla\mu$ where $M=D/kT$ is the particle
mobility. Mass conservation finally gives for the domain growth law:
\begin{equation}\label{Huse}
 \frac{dR}{dt}= \frac{j}{c}=\frac{\gamma D}{R^2\Delta c kT}
\end{equation}
from which we recover directly the scaling law (\ref{scaling}). 

In the following we shall compare our experimental results within this
framework. The different durations of heat treatments allow us to
study the temporal scaling of the coarsening regime and beyond, to
estimate the diffusion constant. In a similar way, data from different
temperatures of heat treatment, combined with the Arrhenian scaling of
the mobility, $M=M_0\exp(-E_{in}/kT)$ provides a rough estimate of the
corresponding activation energy.

\subsection{Effect of surface}
\begin{figure}
\begin{center}
\includegraphics[width=6cm]{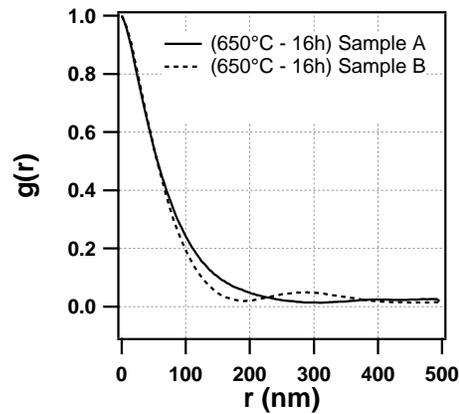}
\caption{Radial mean value of the auto-correlation function $g(r)$
{\it vs} distance $r$ (in nm) for borosilicate glass samples annealed
16 hours at 650$^\circ$C. Samples A have been cut and polished after
annealing treatment and conversely samples B have been cut and
polished before annealing treatment}
\label{surf-bulk}
\end{center}
\end{figure}

We summarize on Fig. \ref{surf-bulk} the height autocorrelation
functions obtained after a heat treatment of 16 hours at
$650^\circ{\mathrm C}$. Two types of preparation were used. As
described above, the heat treatment was applied either before (series
A) or after (series B) cutting and polishing the samples. This allows
to test the effect of a surface on the coarsening process. As shown on
Fig. \ref{surf-bulk} the correlation functions obtained on the two
series of samples are almost identical. The presence of a surface thus
does not seem to influence the coarsening and surface measurements can
be considered as representative of the bulk morphology. The results
presented in the following have been obtained by statistical averaging
over the entire set of samples.

\subsection{Effect of annealing time} \label{duree}

\begin{figure}
\begin{center}
\includegraphics{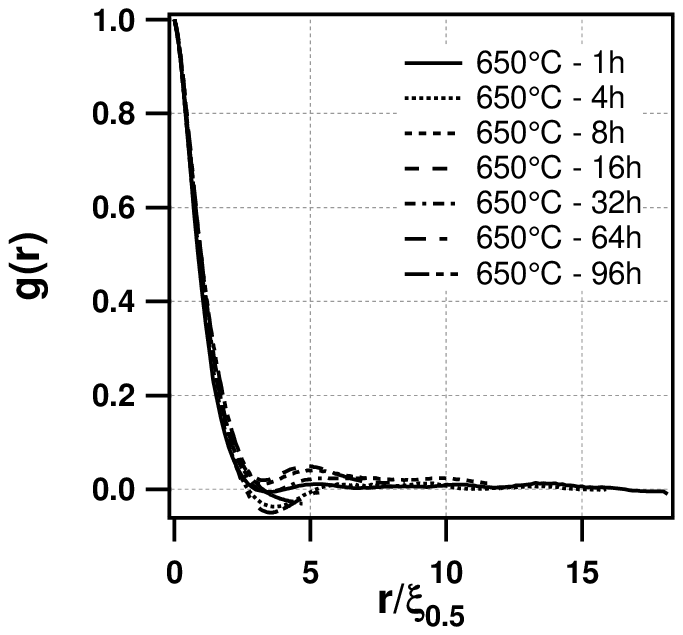}\includegraphics{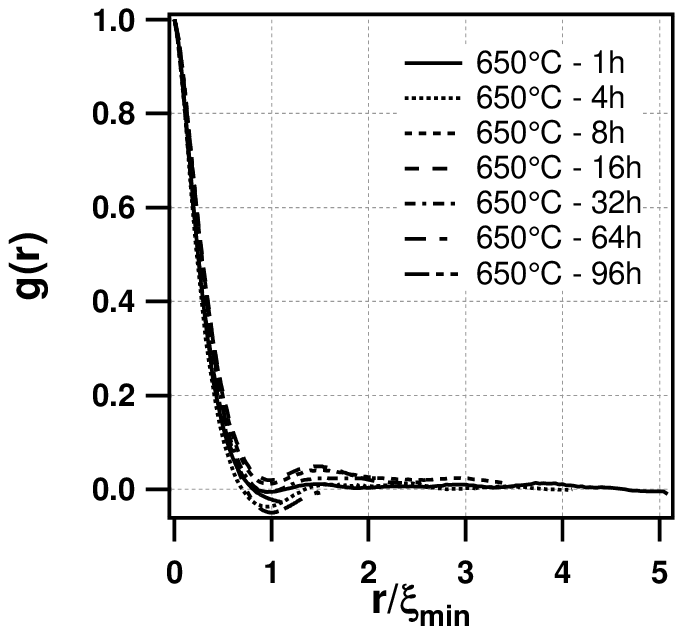}
\caption {Radial mean value of the auto-correlation function (g(r)) as a function of the normalized distance ($r/\xi$) for borosilicate glass samples after thermal treatment during different durations at 650$^\circ$C : (left) r is normalized with the correlation length $\xi_{0.5}$ ($g(\xi_{0.5})=0.5$) and (right) with the correlation length $\xi_{min}$ ($g(\xi_{min})=min(g(r))$).}
\label{tps_norm}
\end{center}
\end{figure}

Let us note first that all samples described here were submitted to a
primary annealing thermal treatment of 1h at 630$^\circ$C. The latter
is intended to relax thermal stresses but also induces phase
separation in the present case. Additional thermal treatments thus only
tend to coarsen the existing phase separated domains. As an
illustration,  we plotted on Fig ~\ref{tps_norm} the auto-correlation
functions $g(r)$ of all samples after rescaling $r/\xi$ both for
$\xi=\xi_{0.5}$ and $\xi=\xi_{min}$. As expected in the coarsening
stage, we observe that all correlation functions fall onto a master
curve, indicating that, regarding the morphology of phase separation,
the only changing parameter is a characteristic length scale.

We report on Fig.~\ref{tps_lc} the time evolution of both correlation
lengths $\xi_{0.5}$ and $\xi_{min}$ in function of the duration of
thermal treatment $t$ at temperature 650$^\circ$C. In this plot, we
use the cubic root of time for the abscissa in order to reveal the
expected scaling. In these coordinates we get a very good linear
behavior over the entire time range experimentally studied (from 1h to
96h). The time evolution of the size of phase separated domains
evaluated by AFM thus obeys very well the predicted scaling for
coarsening. Extracting the leading coefficients by performing linear
fits we obtain $\gamma D \sim 10^{-18}$ where we took $v\sim
10^{-29}\mathrm{m}^{3}$ for a molecular volume. Assuming for the
surface tension a value of order $\gamma \sim
10^{-2}\mathrm{J.m}^{-2}$ this gives us for the value of the
interdiffusion coefficient at 650$^\circ$
$D\sim10^{-16}\mathrm{cm.s}^{-1}$, which is consistent with
literature\cite{MazurinBook84}.

\subsection{Effect of annealing temperature} \label{temperature}

The same analysis is repeated to analyze the evolution of the
correlation lengths $\xi_{0.5}$ and $\xi_{min}$ for 16h long heat
treatments at temperatures varying from $625^\circ\mathrm{C}$ to
$680^\circ\mathrm{C}$. We now test the Arrhenius dependence of the
interdiffusion coefficient. The results are reported on Fig
~\ref{temp_lc} in coordinates $\log \xi$ {\it vs} $1/T$. We obtain
again a reasonable linear behavior indicating that the Arrhenius
scaling gives a good approximation of the dependence of the
interdiffusion coefficient with temperature. 

The values of $c$ and $d$ parameters obtained {\it via} the linear fit
$\ln\xi=c-{d}/{T}$ are given in table~\ref{regtemp} with their
associated standard deviations both for $\xi=\xi_{0.5}$ and for
$\xi=\xi_{min}$. Using $d=E_{in}/3RT$ (the factor $3$ coming from the
exponent $1/3$ in Eq. (\ref{scaling})) we can give a rough estimate of
the activation energy $E_{in}=135 \pm 45 \mathrm{kJ.mol}^{-1}$. This
value seems to be consistent with typical results from literature: by
electrical resistivity measurements in ion exchanged Corning 7740
borosilicate glasses, Garfinkel\cite{Garfinkel-PCG70} obtained
activation energies in the range $80-120 \mathrm{kJ.mol}^{-1}$;
activation energies measured for the diffusion of sodium in silicate
networks are also in the same range\cite{DoremusBook73}.

\section{Conclusion}

Performing selective etching on the surface of phase separated alkali
borosilicate glasses, we revealed the phase separated domain by AFM. A
quantitative statistical analysis of AFM data allowed us to extract
the characteristic correlation lengths of the coarsening stage of
phase separation. A systematic study of the dependence on time and
temperature of the phase separated domains was performed. We recover
the expected cubic root time scaling for the domain size temporal
evolution and we could extract an estimate of the interdiffusivity
coefficient. The dependence of the latter on temperature is consistent
with an Arrhenius behavior, allowing to give a rough estimate of the
activation energy for interdiffusion. The numerical vaues obtained for
these two parameters are moreover consistent with literature data.
This study performed on a very simple phase separated glass thus shows
that, beyond its performance in imaging, AFM analysis can be used in a
more quantitative way to characterize phase separation, and be
considered as an alternative or complementary technique to visible or
X-ray scattering traditionally used in that purpose.

\section*{Acknowledgement}
We benefited from the technical help of A. furet and P. Lambremont 
for the glass preparation. We thank M.H. Chopinet, P. Garnier,
S. Papin, S. Roux and T. Sarlat for useful comments and discussions.




\bibliographystyle{unsrt}
\bibliography{BS,vdb,glass-surface}

\begin{thebibliography}{10}

\bibitem{Arribart-00}
H.~Arribart and D.~Abriou.
\newblock Ten years of atomic force microscopy in glass research.
\newblock {\em Ceramics-Silic\'aty}, 44:121--128, 2000.

\bibitem{Raberg-JNCS05}
W.~Raberg, A.H. Ostadrahimi, T.~Kayser, and K.~Wandelt.
\newblock Atomic scale imaging of amorphous silicate glass surfaces by scanning
  force microscopy.
\newblock {\em J. Non-Cryst. Solids}, 351:1089--1096, 2005.

\bibitem{Frischat-JNCS04}
G.H. Frischat, J.-F. Poggemann, and G.~Heide.
\newblock Nanostructure and and atomic structure of glass seen by atomic force
  microscopy.
\newblock {\em J. Non-Cryst. Solids}, 345-346:197--202, 2004.

\bibitem{Daguier-PRL97}
P.~Daguier, B.~Nghi{\^e}m, E.~Bouchaud, and F.~Creuzet.
\newblock Pinning and depinning of crack fronts in heterogeneous materials.
\newblock {\em Phys. Rev. Lett.}, 78:1062, 1997.

\bibitem{SLSV-Preprint06}
T.~Sarlat, A.~Lelarge, E.~S{\o}nderg{\aa}rd, and D.~Vandembroucq.
\newblock Frozen capillary waves on glass surfaces: an afm study.
\newblock {\em Preprint}, XX:YY--YY, 2006.

\bibitem{kim-jms99}
Hwan Kwang~Lee Je~Young~Kim and Sung~Chul Kim.
\newblock Surface structure and phase separation mechanism of polysulfone
  membranes by atomic force microscopy.
\newblock {\em Journal of Membrane Science}, 163(2):159--166, 1999.

\bibitem{zdravkova-jcg05}
A.N. Zdravkova, J.P.J.M. van~der Eerden, and M.M.E. Snele1029-e1033.
\newblock Phase behaviour in supported mixed monolayers of alkanols,
  investigated by afm.
\newblock {\em Journal of Crystal Growth}, 275(1-2):e1029--e1033, 2005.

\bibitem{Cyganik-ss02}
P.~Cyganik, A.~Budkowski, J.~Raczkowska, and Z.~Postawa.
\newblock Afm/lfm surface studies of a ternary polymer blend cast on substrates
  covered by a self-assembled monolayer.
\newblock {\em Surface Science}, 507-510:700--706, 2002.

\bibitem{rasmont-csb00}
A.~Rasmont, Ph. Leclère, C.~Doneux, G.~Lambin, J.~D. Tong, R.~Jérôme, J.~L.
  Brédas, and R.~Lazzaroni.
\newblock Microphase separation at the surface of block copolymers, as studied
  with atomic force microscopy.
\newblock {\em Colloids and Surfaces B: Biointerfaces}, 19(4):381--395, 2000.

\bibitem{ichii-ass03}
T.~Ichii, T.~Fukuma, K.~Kobayashi, H.~Yamada, and K.~Matsushige.
\newblock Phase-separated alkanethiol self-assembled monolayers investigated by
  non-contact afm.
\newblock {\em Appl. Surf. Sci.}, 210(1-2):99--104, 2003.

\bibitem{bergues-v01}
B.~Bergues, J.~Lekki, A.~Budkowski, P.~Cyganik, M.~Lekka, A.~Bernasik, J.~Rysz,
  and Z.~Postawa.
\newblock Phase decomposition in polymer blend films cast on homogeneous
  substrates modified by self-assembled monolayers.
\newblock {\em Vacuum}, 63(1-2):297--305, 2001.

\bibitem{MazurinBook84}
O.V. Mazurin and E.A. Porai-Koshits, editors.
\newblock {\em Phase separation in glass}.
\newblock North-Holland, Amsterdam, 1984.

\bibitem{tanaka-jncs84}
H.~Tanaka, T.~Yazawa, K.~Eguchi, H.~Nagasawa, N.~Matsuda, and T.~Einishi.
\newblock Precipitation of colloidal silica and pore size distribution in high
  silica porous glass.
\newblock {\em J. Non-Cryt. Solids}, 65(2-3):301--309, 1984.

\bibitem{polizzi-jncs98}
S.~Polizzi, P.~Riello, G.~Fagherazzi, and N.~F. Borrelli.
\newblock The microstructure of borosilicate glasses containing elongated and
  oriented phase-separated crystalline particles.
\newblock {\em J. Non-Cryt. Solids}, 232-234:147--154, 1998.

\bibitem{polizzi-jac97}
S.~Polizzi, P.~Riello, G.~Fagherazzi, M.~Bark, and N.F. Borrelli.
\newblock Two-dimensional small-angle x-ray scattering investigation of
  stretched borosilicate glasses.
\newblock {\em Journal-of-Applied-Crystallography}, 30:487--494, 1997.

\bibitem{tomozawa-jacs80}
M.~Tomozawa and T.~Takamori.
\newblock Scattering study of microstructurally birefringent glasses.
\newblock {\em J. Am. Ceram. Soc.}, 63(5-6):276--280, 1980.

\bibitem{Danforth-jncs83}
S.C. Danforth and J.S. Haggerty.
\newblock Microstructural characterization of graded-index antireflective
  films.
\newblock {\em J. Am. Ceram. Soc.}, 66(1):C6--C8, 1983.

\bibitem{Elmer-jacs70}
T.H. Elmer, M.E. Nordberg, G.B. Carrier, and E.J. Korda.
\newblock Phase separation in borosilicate glasses as seen by electron
  microscopy and scanning electron microscopy.
\newblock {\em J. Am. Ceram. Soc.}, 53(4):171--175, 1970.

\bibitem{polizzi-mmm97}
S.~Polizzi, A.~Armigliato, P.~Riello, N.F. Borrelli, and G.~Fagherazzi.
\newblock Redrawn phase-separated borosilicate glasses: a tem investigation.
\newblock {\em Microscopy Microanalysis Microstructure}, 8(3):157--165, 1997.

\bibitem{kato-jacs01}
Y.~Kato, H.~Yamazaki, and M.~Tomozawa.
\newblock Detection of phase separation by ftir in a liquid-crystal-display
  substrate aluminoborosilicate glass.
\newblock {\em J. Am. Ceram. Soc.}, 84(9):2111--2116, 2001.

\bibitem{koni-jncs76}
W.L. Konijnendijk and J.M. Stevels.
\newblock The structure of borosilicate glasses studied by raman scattering.
\newblock {\em J. Non-Cryt. Solids}, 20(2):193--224, 1976.

\bibitem{White-jncs82}
W.B. White.
\newblock Investigation of phase separation by raman spectroscopy.
\newblock {\em J. Non-Cryt. Solids}, 49(1-3):321--329, 1982.

\bibitem{sigoli-jncs01}
F.A. Sigoli, Y.~Kawano, M.R. Davolos, and M.Jr Jafelicci.
\newblock Phase separation in pyrex glass by hydrothermal treatment: evidence
  from micro-raman spectroscopy.
\newblock {\em J. Non-Cryt. Solids}, 284:49--54, 2001.

\bibitem{Malik-PRL98}
A.~Malik, A.R. Sandy, L.B. Lurio, G.B. Stephenson, S.G.J. Mochrie, I.~McNulty,
  and M.~Sutton.
\newblock Coherent x-ray study of fluctuations during domain coarsening.
\newblock {\em Phys. Rev. Lett.}, 81:5832--5835, 1998.

\bibitem{BalluffiBook05}
R.W Balluffi, S.M. Allen, and W.C. Carter, editors.
\newblock {\em Kinetics of materials}.
\newblock Wiley, Hoboken, 2005.

\bibitem{Huse-PRB86}
D.A. Huse.
\newblock Corrections to late-stage behavior in spinodal decomposition:
  Lifshitz-slyozov scaling and monte carlo simulations.
\newblock {\em Phys. Rev. B}, 34:7845, 1986.

\bibitem{Garfinkel-PCG70}
H.M. Garfinkel.
\newblock Ion exchange properties of borosilicate glass membranes.
\newblock {\em Phys. Chem. Glass.}, 11:151--158, 1970.

\bibitem{DoremusBook73}
R.H. Doremus.
\newblock {\em Glass science}.
\newblock Wiley-Interscience, New-York, 1973.

\end{thebibliography}
\addcontentsline{toc}{section}{Références}






\newpage

\begin{table}
  \centering
   \begin{tabular}{|c|c|c|}\hline
  \textbf{Duration}& $\xi_{0.5}$ (nm) & $\xi_{min}$ (nm) \\ \hline
  1 hour & 27 & 97 \\
  4 hours & 31 & 121 \\
  8 hours & 43 & 144 \\
  16 hours & 58 & 195 \\
  32 hours & 66 & 238 \\
  64 hours & 94 & 332 \\
  96 hours & 105 & 422 \\ \hline
    \end{tabular}
  \caption{Correlation length ($\xi_{0.5}$ and $\xi_{min}$) {\it vs} duration of thermal treatment at 650$^\circ$C.}\label{tabtps}
\end{table}

\begin{table}
 \begin{center}
\begin{tabular}{|c|c|c|}\hline
  \textbf{Temperature}& $\xi_{0.5}$ (nm) & $\xi_{min}$ (nm) \\ \hline
  625$^\circ$C & 47 & 148 \\
  650$^\circ$C & 58 & 195 \\
  660$^\circ$C & 51 & 160 \\
  670$^\circ$C & 62 & 191 \\
  675$^\circ$C & 63 & 207 \\
  680$^\circ$C & 66 & 230 \\ \hline
\end{tabular}
\end{center}
\caption{Correlation length ($\xi_{0.5}$ and $\xi_{min}$)  {\it vs} temperature for 16 hours thermal treatment.}\label{tabtemp}
\end{table}

\begin{table}
  \centering
   \begin{tabular}{|c|c|c|}\hline
  $\ln\xi=c-\frac{d}{T}$ &  c &  d \\ \hline
  $\xi_{0.5}$ & $-11.34\pm 1.54$ & $(4.9\pm1.4)\cdot10^{3}$ \\
  $\xi_{min}$ &$-9.43\pm 2.13$& $(5.6\pm1.9)\cdot10^{3}$ \\ \hline
    \end{tabular}
 \caption{Parameters from linear curve fitting of $\ln\xi_{0.5}$ and $\ln\xi_{min}$ as a function of the inverse of absolute temperatures $(1/T)$ for 16 hours thermal treatments.}\label{regtemp}
\end{table}

\end{document}